\documentclass[useAMS,usenatbib]{mn2e}
\usepackage{amsmath,natbib,lscape,graphicx,natbib,amssymb,times,multirow,epsfig,longtable,color,rotating,placeins,url}

\title[A wide substellar companion to $\zeta$ Delphini]{The VAST
  Survey -- IV. A wide brown dwarf companion to the A3V star $\zeta$ Delphini\thanks{Based on observations obtained at the Canada-France-Hawaii Telescope (CFHT) under programme ID 2009BC06, observations obtained at the Gemini Observatory under programme IDs  GN-2008B-Q-119, GN-2010A-Q-75, GN-2013A-Q-96 and GN-2013B-DD-4, and observations obtained at the MMT Observatory, a joint facility of the University of Arizona and the Smithsonian Institution.}}

\author[De Rosa et al.]
{R. J. De Rosa$^{1,2}$\thanks{E-mail: derosa@berkeley.edu}, J. Patience$^1$, K. Ward-Duong$^1$, A. Vigan$^3$, C. Marois$^4$, I. Song$^5$,\newauthor
  B. Macintosh$^{6,7}$, J. R. Graham$^8$, R. Doyon$^9$,  M. S. Bessell$^{10}$, O. Lai$^{11,12}$, D. W. McCarthy$^{13}$ \newauthor \& C. Kulesa$^{13}$\\
$^1$ School of Earth and Space Exploration, Arizona State University,
PO Box 871404, Tempe, AZ 85287, USA\\
$^2$ School of Physics, College of Engineering, Mathematics and Physical Sciences, University of Exeter, Stocker Road, Exeter, EX4 4QL,
UK\\
$^3$ Aix Marseille Universit\'{e}, CNRS, LAM (Laboratoire d'Astrophysique de Marseille) UMR 7326, 13388 Marseille, France\\ 
$^4$ NRC Herzberg Astronomy and Astrophysics, 5071 West Saanich Road, Victoria, BC, V9E 2E7, Canada\\
$^5$ Physics and Astronomy, University of Georgia, 240 Physics, Athens, GA 30602,
USA\\
$^6$ Kavli Institute for Particle Astrophysics and Cosmology, Stanford University, Stanford, CA 94305, USA\\
$^7$ Institute of Geophysics and Planetary Physics, Lawrence Livermore
National Laboratory, 7000 East Ave, Livermore, CA 94550,
USA\\
$^8$ Department of Astronomy, University of California at Berkeley,
Berkeley, CA 94720, USA\\
$^9$ D\'{e}pt de Physique, Universit\'{e} de Montr\'{e}al, C.P.
6128, Succ. Centre-Ville, Montr\'{e}al, QC,
H3C 3J7, Canada\\
$^{10}$ Research School of Astronomy and Astrophysics, Mount Stromlo Observatory, The Australian National University, ACT 2611,
Australia\\
$^{11}$ Gemini Observatory, 670 N. A'ohoku Place, Hilo, HI 96720, USA\\
$^{12}$ National Astronomical Observatory of Japan, 650 North A'ohoku Place, Hilo, HI 96720, USA\\
$^{13}$ Steward Observatory, University of Arizona, 933 N. Cherry Ave, Tucson, AZ 85721, USA}

\date{-}
\pagerange{\pageref{firstpage}--\pageref{lastpage}} \pubyear{2014}
\begin{document}
\label{firstpage}

\maketitle

\begin{abstract}
We report the discovery of a wide comoving substellar companion to the nearby ($D=67.5\pm1.1$~pc) A3V star $\zeta$~Delphini based on imaging and follow-up spectroscopic observations obtained during the course of our Volume-limited A-Star (VAST) multiplicity survey. $\zeta$~Del was observed over a five-year baseline with adaptive optics, revealing the presence of a previously-unresolved companion with a proper motion consistent with that of the A-type primary. The age of the $\zeta$~Del system was estimated as $525\pm125$~Myr based on the position of the primary on the colour-magnitude and temperature-luminosity diagrams. Using intermediate-resolution near-infrared spectroscopy, the spectrum of $\zeta$~Del~B is shown to be consistent with a mid-L dwarf (L$5\pm2$), at a temperature of $1650\pm200$~K. Combining the measured near-infrared magnitude of $\zeta$~Del~B with the estimated temperature leads to a model-dependent mass estimate of $50\pm15$~$M_{\rm Jup}$, corresponding to a mass ratio of $q=0.019\pm0.006$. At a projected separation of $910\pm14$ au, $\zeta$~Del~B is among the most widely-separated and extreme-mass ratio substellar companions to a main-sequence star resolved to-date, providing a rare empirical constraint of the formation of low-mass ratio companions at extremely wide separations.
\end{abstract}

\begin{keywords} techniques: high angular resolution - techniques: spectroscopic - binaries: visual - brown dwarfs - stars: early-type - stars: individual: $\zeta$ Del
\end{keywords}

\section{Introduction}
Since the discovery of the brown dwarf GJ~229~B \citep{Nakajima:1995bb}, over a hundred examples of brown dwarf companions to main-sequence stars have been catalogued (e.g. \citealp{Bird:2010th}). Recent surveys designed to characterize the frequency of substellar companions to nearby stars have found a significant deficit of brown dwarf companions relative to both planetary-mass and stellar companions \citep{Grether:2006ek,Lafreniere:2007cv,Vigan:2012jm,Nielsen:2013jy}. The dearth of brown dwarf companions is most striking when compared to known planet-hosting systems: an order of magnitude more planetary-mass companions have been discovered to date, despite being more technically challenging to detect. Understanding the frequency and properties of companions intermediate to stellar and planetary-mass companions, and determining the shape of the companion mass ratio distribution, will allow for the formation of such companions to be placed in the context of both binary star and planet formation theories (e.g. \citealp{Chabrier:2014up}). 

At wide angular separations, typically greater than an arcsecond, substellar companions are readily accessible to spectroscopic follow-up observations to characterize their atmospheres (e.g. \citealp{Schultz:1998ge,Leggett:2008kq,Janson:2010db,Chilcote:2014vu}). These wide substellar companions represent important benchmark objects with which atmospheric and evolutionary models can be tested. By assuming coevality of the companion with the main-sequence primary, the observed degeneracies between age, mass, and luminosity for these objects can be broken (e.g. \citealp{Dupuy:2009jq,Kasper:2009dc,King:2010iu,Deacon:2012eg}). Considerable effort has been made in identifying wide substellar companions to main-sequence stars, with studies utilizing both all-sky photometric surveys (e.g. \citealp{Gizis:2001ew,Pinfield:2006bp,Faherty:2010gt}) and dedicated multi-epoch surveys (e.g. \citealp{Deacon:2014tf}) discovering numerous examples, predominantly with field-age solar-type primaries.

In this paper we report the astrometric and spectroscopic confirmation of a newly-discovered wide substellar companion to the A3V star $\zeta$~Delphini (hereafter $\zeta$~Del), observed over a five-year baseline during the course of our Volume-limited A-Star (VAST) multiplicity survey \citep{DeRosa:2011ci,DeRosa:2012gq,DeRosa:2014db}. The companion to $\zeta$~Del joins the small list of known brown dwarf companions to early-type ($<$F0) stars: HR~7329~B \citep{Lowrance:2000ic}, HD~100546~B \citep{Acke:2006gn,Mulders:2013kr}, HD~180777~B \citep{Galland:2006ie}, HR~6037~B (\citealp{Huelamo:2010cx}, discovered to be a binary brown dwarf by \citealp{Nielsen:2013jy}), HIP~78530~B \citep{Lafreniere:2011dh}, HD 1160~BaBb \citep{Nielsen:2012jk}, and $\kappa$~Andromedae~b \citep{Carson:2013fw,Bonnefoy:2014dx}. The large majority of these substellar companions were discovered using direct-imaging techniques; precision radial velocity searches for low-mass companions to A-type stars are complicated by the limited number of metallic lines, and their broadening caused by rapid stellar rotation. With the relatively low yield of brown dwarf companions in recent large-scale surveys of nearby, young ($<1$~Gyr) stars, the detection of an individual companion still represents a significant advancement in our understanding of these objects.

\section{Properties of $\zeta$ Del A}
\begin{figure}
\resizebox{\hsize}{!}{{\includegraphics{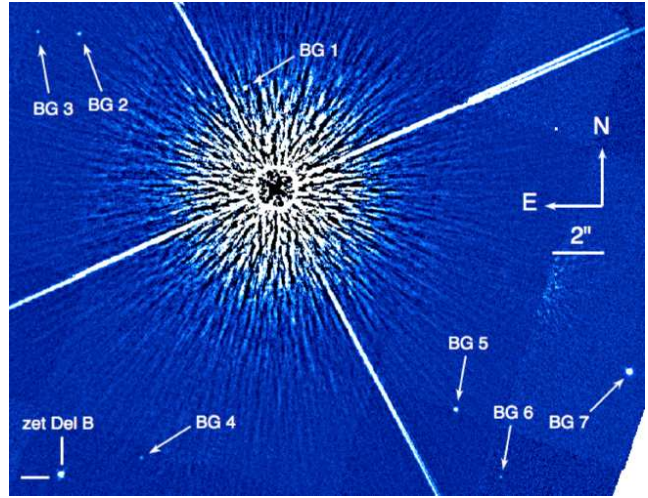}}} 
\caption{The Gemini/NIRI observation of the $\zeta$~Del system obtained on 2010 June 8 showing the location of the heavily saturated $\zeta$~Del~A, the substellar companion $\zeta$~Del~B (indicated) and the seven background objects used in the astrometric analysis (indicated by the arrows). The image has been processed through a median filter to reduce the significant amount of scattered light from $\zeta$~Del~A. The orientation and angular scale are given for reference.}
\label{fig:geminiAO}
\end{figure}
\begin{figure}
\resizebox{\hsize}{!}{{\includegraphics{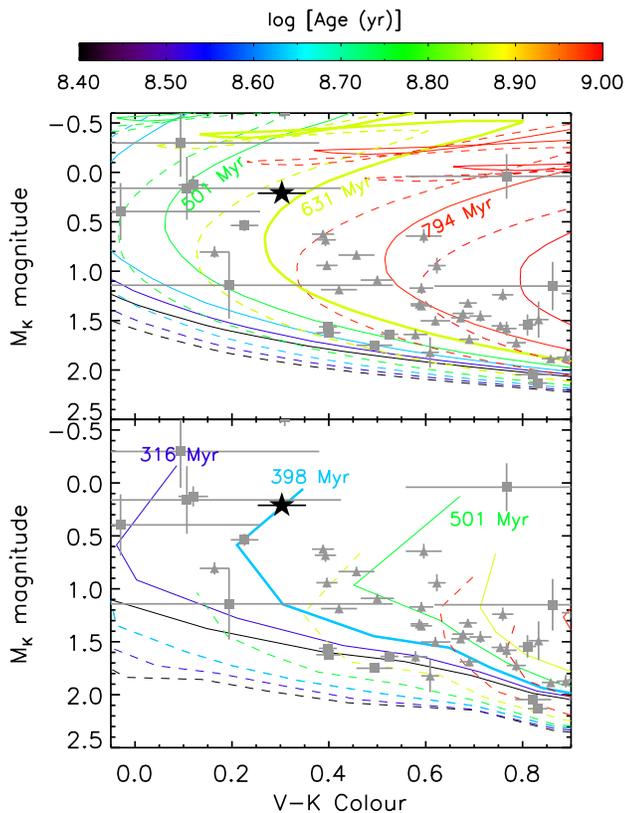}}} 
\caption{The position of $\zeta$~Del~A (filled black star) on the CMDs compared with the stellar evolutionary models of \citealp{Bressan:2012bx} (top panel) and \citealp{Siess:2000tk} (bottom panel). The solid lines represent isochrones with $Z=0.02$, coloured according to age. For the \citet{Bressan:2012bx} grid, the dashed lines are isochrones with $Z=0.0135$, based on the assumption that [$M$/H]$\sim$[Fe/H]$=-0.05$~dex \citep{Erspamer:2003hz} and $Z_{\odot}=0.0152$. For the \citet{Siess:2000tk} grid, the dashed lines are for isochrones with $Z=0.01$. Isochrones are plotted with ages in the range $\log(t)=8.4-9.0$, with a spacing of $\Delta\log(t)=0.1$. The \citet{Bressan:2012bx} grid suggests a significantly older age for $\zeta$~Del~A of ${\sim}630$~Myr, compared with ${\sim}400$~Myr from the \citet{Siess:2000tk} grid. We therefore assign an age of $525\pm125$~Myr, encompassing both of these estimates. Early-type members of the Ursa Majoris moving group (grey squares - $300-500$~Myr; \citealp{King:2003df,Zuckerman:2004ex}), and the Hyades open cluster (grey triangles - 625~Myr; \citealp{Perryman:1998vj}) are plotted for reference, with the two clusters bracketing the location of $\zeta$~Del~A, consistent with the assigned age of $525\pm125$~Myr. The large photometric uncertainty for a subset of the Ursa Majoris and Hyades members is due to their saturation within the 2MASS {\it K}-band images.}
\label{fig:isochrones}
\end{figure}

Observed as a part of our VAST multiplicity survey, $\zeta$~Del~A is an A3V star \citep{Slettebak:1954bc} located at a distance of $67.5\pm1.1$~pc from the Sun \citep{vanLeeuwen:2007dc}, with $T_{\rm eff}=8336$~K, $\log (g)=3.72$~dex and a slightly subsolar metallicity of [Fe/H]$=-0.05$ (Table \ref{tab:zetDel}; \citealp{Erspamer:2003hz}). Previous speckle interferometry measurements, and the adaptive optics (AO) measurements presented here, exclude the presence of a bright ($\Delta V \lesssim 3$) stellar companion beyond $0.03$~arcsec (2~au; \citealp{Hartkopf:1984df}), and any stellar companion to the bottom of the main sequence beyond ${\sim}0.9$~arcsec (60~au). Eight substellar candidates were identified within the AO imaging presented in this study within a separation of ${\sim}15$~arcsec from $\zeta$~Del~A, as shown in Fig. \ref{fig:geminiAO}. Searches for stellar companions beyond the field of view of the AO images have yielded a null result \citep{Shaya:2010dt,DeRosa:2014db}.  $\zeta$~Del~A is listed as having either a spectroscopic companion or radial velocity variations by \citet{Hartkopf:1984df}, although no further spectroscopic observations could be found in the literature confirming or rejecting these variations. \citet{Hartkopf:1984df} also list the star as potentially being photometrically variable, though subsequent {\it Hipparcos} measurements exclude variability amplitudes greater than 0.01~mag \citep{Adelman:2001iw}.

$\zeta$~Del~A is not a known member of any stellar moving group or association, and is not well-suited for age determination through either gyrochronology or chromospheric indicators due to the lack of surface convection zones and diminishing chromospheric activity observed for A-type stars (e.g. \citealp{Barnes:2003ga,Mamajek:2008jz}). An age can therefore only be estimated through a comparison of the observed and derived parameters with theoretical stellar evolution models. Two grids of models were used to estimate the age \citep{Siess:2000tk,Bressan:2012bx}. The position of the star on the temperature-gravity and the colour-magnitude diagrams (CMDs) was used to estimate the age, with the CMD shown in Fig. \ref{fig:isochrones}. The temperature and surface gravity of $\zeta$~Del~A were estimated from high-resolution optical spectroscopy \citep{Erspamer:2003hz}. Optical and near-infrared absolute magnitudes colours were computed from measurements within the {\it Tycho2} \citep{Hog:2000wk} and 2MASS \citep{Skrutskie:2006hla} catalogues, respectively. The bolometric luminosity of $\zeta$~Del~A was then estimated to be $\log\left(L/{\rm L}_{\odot}\right)=1.687\pm0.015$ based on the observed {\it B}$-${\it V} colour, a bolometric correction \citep{Flower:1996ima}, and the solar bolometric magnitude of $M_{\rm bol,\odot}=4.74$ \citep{Drilling:2000vo}. These observed and derived stellar properties are summarized in Table \ref{tab:zetDel}. The age for $\zeta$~Del~A was estimated to be ${\sim}400$~Myr using the \citet{Siess:2000tk} grid, and ${\sim}630$~Myr using the \citet{Bressan:2012bx} grid. While neither of these grids take into account the effects of stellar rotation (e.g. \citealp{Ekstrom:2012ke}), the effect on the age estimate is most significant for stars near the zero age main sequence, where a small change in the observed luminosity and temperature of a star can cause a large change in the age derived from its position on the CMD relative to theoretical isochrones \citep{Collins:1985vd}. As $\zeta$~Del~A has evolved away from the zero age main sequence, the effect of rotation on the age estimate is likely to be small.

Taking both of these age estimates into account, we assigned an age of $525\pm125$~Myr to $\zeta$~Del~A, consistent with its location on the CMD intermediate to the A-type stars within the 300--500~Myr Ursa Majoris moving group \citep{King:2003df,Zuckerman:2004ex} and 625~Myr Hyades open cluster \citep{Perryman:1998vj}, as shown in Fig. \ref{fig:isochrones}. The $125$~Myr uncertainty on the age corresponds to the range of ages estimated from the two model grids, as opposed to a formal statistical uncertainty. Another age diagnostic for early-type stars is the presence of a circumstellar debris disc (e.g. \citealp{Gaspar:2013is}). Based on the expected photospheric flux of $\zeta$~Del~A at 22~$\micron$, and the reported {\it W4} magnitude within the {\it Wide-field Infrared Survey Explorer} catalogue \citep{Wright:2010in}, we find no evidence of an infrared excess indicative of the presence of warm circumstellar material, consistent with the age estimated for $\zeta$~Del~A relative to the bulk of debris disc-hosting early-type stars \citep{Rieke:2005hv,Su:2006ce}. No longer-wavelength {\it Spitzer} or {\it Herschel} measurements were found within the literature. The mass of $\zeta$~Del~A was estimated as $2.5\pm0.2$~M$_{\odot}$ based on the position of the primary on both the temperature-surface gravity and CMDs relative to mass tracks within each model grid. 

\section{AO imaging observations and results}
\begin{table*}
\caption{Astrometry and photometry derived from AO observations of the $\zeta$ Del system}
\begin{tabular}{ccccccccc}
Instrument&Date&MJD&Filter&Plate scale&True north&Separation&Position angle&$\Delta${\it K}\\
&&&&($10^{-3}$ arcsec px$^{-1}$)&(deg)&(arcsec)&(deg)&(mag)\\
\hline
\hline       
Gemini/NIRI & 2008-09-08 & 54717.2398 & {\it K$^{\prime}$} & $21.27\pm0.12$ & $ 0.51\pm0.30$ & $13.51\pm0.08$ & $143.35\pm0.30$ & --\\
CFHT/KIR    & 2009-08-31 & 55074.4203 & {\it K$^{\prime}$} & $34.75\pm0.09$ & $-2.38\pm0.13$ & $13.47\pm0.04$ & $143.12\pm0.14$ & $10.83\pm0.27$\\
Gemini/NIRI & 2010-06-08 & 55355.6308 & {\it K$^{\prime}$} & $21.27\pm0.12$ & $ 0.51\pm0.30$ & $13.51\pm0.08$ & $143.05\pm0.30$ & --\\
Gemini/NIRI & 2013-05-07 & 56419.6402 & {\it K$^{\prime}$} & $21.27\pm0.12$ & $ 0.51\pm0.30$ & --             & --              & --\\
MMT/ARIES   & 2013-09-18 & 56554.7627 & {\it K$_{\rm S}$}  & $40.57\pm0.18$ & $ 3.22\pm0.24$ & $13.53\pm0.06$ & $143.18\pm0.24$ & $10.99\pm0.16$\\
\hline
\label{tab:ao_obs}
\end{tabular}
\end{table*}
\subsection{Observations and data analysis}
\begin{figure}
\resizebox{\hsize}{!}{{\includegraphics{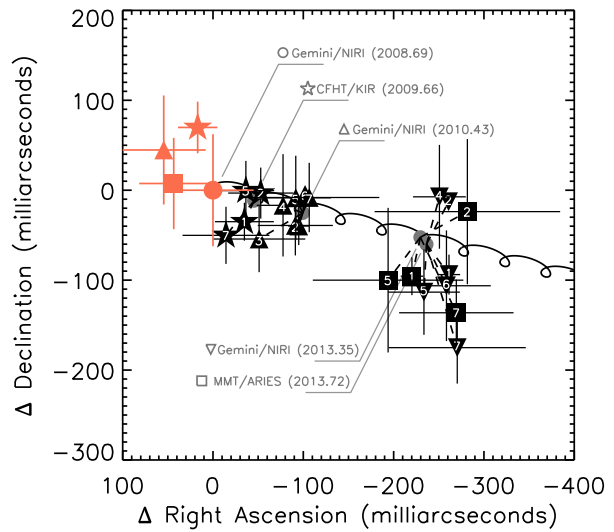}}} 
\caption{The astrometric motion of $\zeta$~Del~B (red filled symbols), and several candidate companions (black numbered symbols), measured relative to $\zeta$~Del~A within the five observational epochs. The measurements within the different epochs are differentiated by the symbol used, with the different objects numbered as in Fig. \ref{fig:geminiAO}. Some objects were not detected within each epoch due to the varying sensitivities and field of views of the instruments used. The expected motion of a background object (black curve) was calculated based on the {\it Hipparcos} proper motion and parallax of $\zeta$~Del~A \citep{vanLeeuwen:2007dc}, with the expected position at each observational epoch indicated (grey filled circles). Dashed lines connect the observed position of each object with the expected location for a background object at the same observational epoch. All subsequent measurements of $\zeta$~Del~B are within the observational uncertainties of the original Gemini/NIRI 2008 measurement, whereas the seven candidate companions exhibit a motion consistent with a stationary background object.}
\label{fig:PM}
\end{figure}
$\zeta$ Del was initially observed on 2008 September 8 using the Near InfraRed Imager and Spectrometer (NIRI; \citealp{Hodapp:2003ko}) in combination with the ALTitude conjugate Adaptive optics for the InfraRed (ALTAIR; \citealp{Herriot:2000wb}) system on the Gemini North telescope. The observing strategy consisted of a sequence of short integrations with the narrow-band Br$\gamma$ filter in which the primary star remains unsaturated, followed by longer exposures in the wide-band {\it K$^{\prime}$} filter to achieve sensitivity to faint stellar and substellar companions. The unsaturated sequence consisted of images obtained at four dither positions on a $256\times256$ subarray of the NIRI detector, each consisting of 400 co-added frames of 0.021~s. The saturated sequence, using the full $1024\times1024$ array, also used a four-point dither pattern, at which two co-added frames of 25~s were obtained. The ALTAIR field lens was used to reduce the effect of isoplanatism.

In order to determine if any of the identified substellar companion candidates were physically bound, additional observations were obtained with Gemini/NIRI in similar configuration in 2010 and 2013. Fig. \ref{fig:geminiAO} shows the 2010 June 6 Gemini/NIRI observations in which the heavily-saturated primary can be seen, in addition to eight candidate substellar companions. Observations were also obtained with the KIR instrument \citep{Doyon:1998vi} on the Canada France Hawaii Telescope (CFHT) in 2009 and with the Arizona Infrared imager and Echelle Spectrograph (ARIES; \citealp{McCarthy:1998wr}) on the MMT in 2013. For the CFHT/KIR observations on 2009 August 31, a sequence of 20 0.5~s exposures were taken using the narrow-band H$_2 (v=1-0)$ filter in a four-point dither pattern on a $512\times512$ subarray of the KIR detector. This was followed by a saturated sequence of 27 60~s exposures taken using the wide-band {\it K$^{\prime}$} filter in a four-point dither pattern on the full $1024\times1024$ array. For the MMT/ARIES observations on 2013 September 18, a sequence of nine 0.9~s exposures were taken using the narrow-band {\it K$_{\rm c, 2.09}$} filter, in combination with a neutral density filter, in a nine-point dither pattern. This was followed by a saturated sequence of nine 4.9~s exposures using the wide-band {\it K$_{\rm S}$} filter in a nine-point dither pattern.

All of the images were processed through the standard near-infrared data reduction process consisting of the following steps: dark frame subtraction, division by a flat-field, bad pixel identification and removal, and sky subtraction. For the Gemini/NIRI observations, the field distortion was corrected based on the information provided on the Gemini website\footnote{http://www.gemini.edu/node/10058}. The position of the primary in the unsaturated images was estimated from a Gaussian fit using the \textsc{GCNTRD} \textsc{IDL} procedure, while in the saturated images the position was determined through a cross-correlation of the diffraction spikes caused by the secondary mirror supports \citep{Lafreniere:2007cv}. For each set of observations, the images were shifted to a common centre using the measured position of the primary, and rotated such that north was up and east to the left based on the astrometric information encoded into the image headers.

\subsection{Astrometry}
\begin{figure}
\resizebox{\hsize}{!}{{\includegraphics{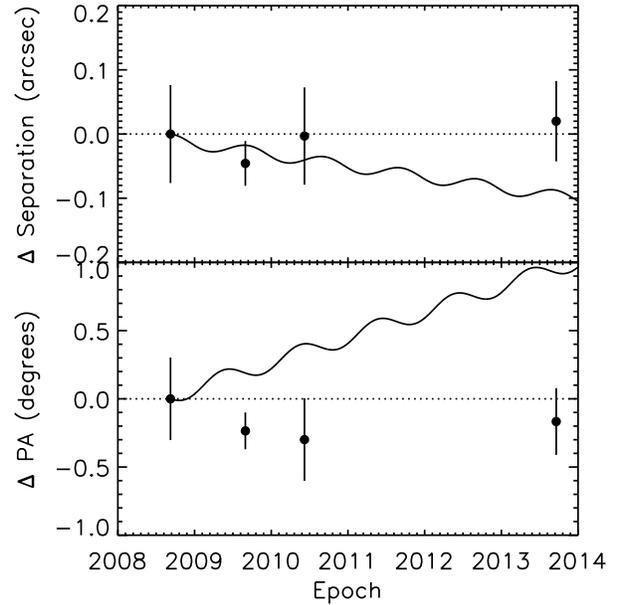}}} 
\caption{The measured changes in the separation (top panel) and position angle (bottom panel) of $\zeta$~Del~B relative to $\zeta$~Del~A (filled points) are inconsistent with the expected change for a stationary background object (solid curve). As $\zeta$~Del~A has a proper motion vector almost tangential to vector between $\zeta$~Del~A and B, the expected motion of a stationary background object relative to $\zeta$~Del~A at the location of $\zeta$~Del~B is dominated by the change in the position angle $\left(d\theta/dt\right)$.}
\label{fig:cpm}
\end{figure}

Based on a visual inspection of the images obtained over the multiple epochs, one of the candidate substellar companions (hereafter $\zeta$~Del~B) was observed to remain stationary with respect to the centroid of $\zeta$~Del~A, with the remaining candidates moving with the magnitude and direction consistent with that of a background object. The relative pixel positions of each object within each of the observations were measured by comparing the centroid location of each object measured using the \textsc{GCNTRD} routine with the position of the saturated primary estimated previously. In order to convert the pixel position into an angular separation and position angle, the plate scale and orientation of each observations was required. Comparing the pixel position of Trapezium cluster members in observations obtained in 2010 with both CFHT/KIR and Gemini/NIRI, and in 2013 with MMT/ARIES, with previous astrometric measurements \citep{McCaughrean:1994id}, enabled a measurement of the plate scale and angle of true north for each instrument, given in Table \ref{tab:ao_obs}.

The pixel scale and orientation of each instrument were used to convert the measured pixel offset between $\zeta$~Del~A and B into an on-sky angular separation and position angle for each observation, which are listed in Table \ref{tab:ao_obs}. This process was repeated for seven additional candidates resolved in the vicinity of $\zeta$~Del~A. The astrometric motion of each candidate was measured relative to the position in the 2008 Gemini/NIRI data set. The relative change in the positions of eight candidate companions from the first epoch, plotted alongside the expected motion of a background objects based on the parallax and proper motion of $\zeta$~Del~A, are shown in Fig. \ref{fig:PM}. Each subsequent measurement of $\zeta$~Del~B was within the uncertainty of the original epoch, consistent with a physically bound companion. In contrast, the seven additional candidates followed the expected motion of a background object (Fig. \ref{fig:PM}). The changes in the separation and position angle of $\zeta$~Del~B measured relative to the 2008 Gemini/NIRI data set are also inconsistent with the expected motion of a background objects, as shown in Fig. \ref{fig:cpm}.

In order to quantify the significance of the astrometric confirmation, the $\chi^2$ statistic is computed based on a comparison of the position of $\zeta$~Del~B with that expected of a background object ($\chi_{\rm BG}^2$), and with that expected of a bound companion with no orbital motion ($\chi_{\rm CPM}^2$). Definitions for $\chi_{\rm BG}^2$ and $\chi_{\rm CPM}^2$ are given in \citet{Nielsen:2012jk}. Comparing the measured position of $\zeta$~Del~B with the expected motion of a background object leads to a $\chi^2_{\rm BG}=39.38$, corresponding to a reduced $\chi^2=6.56$ with 6 degrees of freedom. The measurements are inconsistent with the hypothesis that $\zeta$~Del~B is a stationary background object, with a probability derived from the $\chi^2$ statistic of $P_{\rm BG}\sim 10^{-6}$. Fitting the measurements to the expected constant separation and position angle of a bound companion yields a $\chi^2_{\rm CPM}=6.29$, corresponding to a reduced $\chi^2=1.05$. Based on the value of $\chi_{\rm CPM}^2$, the hypothesis that $\zeta$~Del~B has a fixed separation and position angle can only be accepted at a low confidence level ($P_{\rm CPM}=0.39$), due to the relatively small annular proper motion of $\zeta$~Del~A. Whilst the confidence of the astrometric confirmation is relatively low, additional spectroscopic evidence was obtained to support the hypothesis that $\zeta$~Del~B is a bound companion (Section 4).

\subsection{Photometry}
As the primary was saturated within all of the images in which $\zeta$~Del~B was detected, it was not possible to measure a magnitude difference in an individual image. Instead, the flux from the primary was measured using aperture photometry within the short integrations taken with a narrow-band filter. This flux was then scaled according to the relative exposure times of the unsaturated and saturated exposures, and the relative transmission of the narrow and wide-band filters. While the former quantity is precisely recorded within the header of each image, the latter can be significantly biased by both the temperature at which the filter transmission curve was measured, and the colour of stars being observed due to differences in the effective wavelength of the two filters.

The relative transmission of the CFHT/KIR H$_2(v=1-0)$ and {\it K$^{\prime}$} filters and the MMT/ARIES {\it K$_{\rm c,2.09}$}+ND1 and {\it K$_{\rm S}$} filters were determined empirically using measurements of early-type stars within the Trapezium cluster obtained with each filter combination. Exposure times were selected to minimize the number of saturated stars within each image. A linear fit to the flux for each unsaturated star measured within each pair of narrow and wide-band filters gave an estimate of the relative transmission of 25.4 for the CFHT/KIR observations, and 127.7 for the MMT/ARIES observations. Using these scaling factors, the flux of $\zeta$~Del~A within the unsaturated observations was scaled and compared with that of $\zeta$~Del~B measured within the saturated observations. The magnitude differences measured within the CFHT/KIR and MMT/ARIES observations are listed in Table \ref{tab:ao_obs}. No measurement of the magnitude difference was made for the Gemini/NIRI images due to the limited number of images for which $\zeta$~Del~B was within the field of view; the Gemini/NIRI data were only used for the astrometric analysis.

\section{Spectroscopic Observations and Results}
\begin{figure*}
\resizebox{\hsize}{!}{{\includegraphics{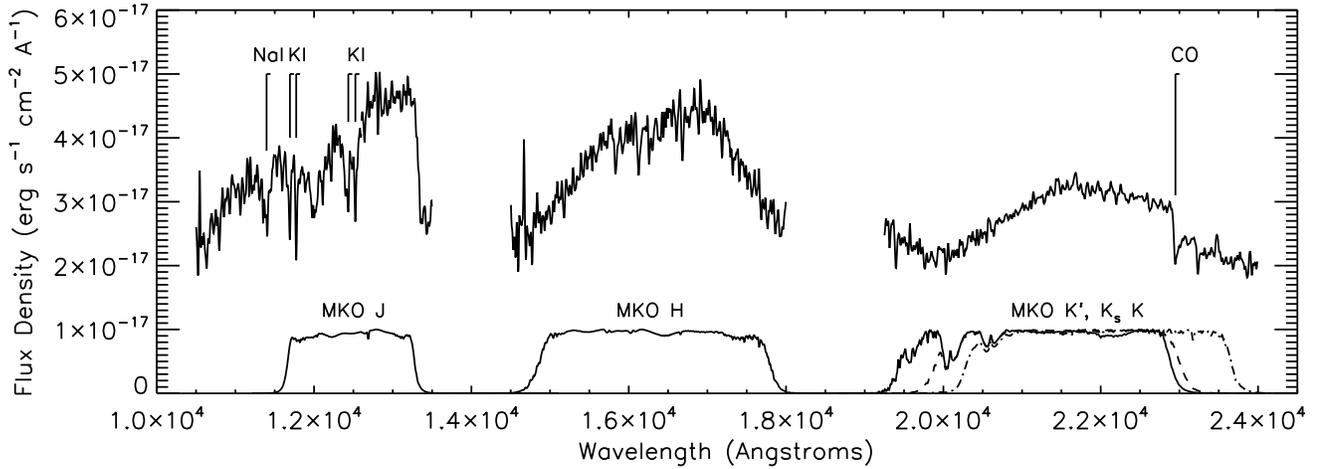}}} 
\caption{The flux-calibrated spectrum of $\zeta$~Del~B obtained on 2013 October 11 with GNIRS. The strong Na {\sc i} line at 1.122~$\micron$, the K {\sc i} doublets at 1.175 and 1.245~$\micron$, and the deep $^{12}$CO bandhead at 2.29~$\micron$ are indicated. The MKO filter transmission curves used to compute the apparent magnitudes for $\zeta$~Del~B in the MKO photometric system are over-plotted \citep{Tokunaga:2002ex}. For clarity, the 2MASS filter transmission curves are not shown.}
\label{fig:GNIRS}
\end{figure*}

An intermediate-resolution ($R\sim1800$) spectrum of $\zeta$~Del~B was obtained with the Gemini Near Infra-Red Spectrograph (GNIRS; \citealp{Elias:2006dh}) mounted on the Gemini North telescope on 2013 October 11 in order to confirm the substellar nature of the object, and reject the possibility of a background star with a similar proper motion to $\zeta$~Del~A. GNIRS was operated in cross-dispersed mode, providing simultaneous coverage to wavelengths in the range $0.9-2.5~\micron$, using the short blue camera with the 32~lines~mm$^{-1}$ grating and the 0.3~arcsec slit. $\zeta$~Del~B was observed in an ABBA pattern, with 300~s per exposure, and a total on-source integration time of 1800~s. The slit was orientated at a position angle of 270~deg, with no contamination from $\zeta$~Del~A seen within either the raw or reduced spectra. An argon lamp and a flat-field lamp were observed to measure the wavelength scale and variations in the detector response. Finally, a spectrum of the B3V star HIP~104320 ($T_{\rm eff}=19000$~K; \citealp{Cox:2000ua}) was obtained, to correct for telluric absorption in the spectrum of $\zeta$~Del~B.

The observations of $\zeta$~Del~B, and the telluric calibrator, were reduced using the GNIRS-specific tools provided within the Gemini \textsc{IRAF} package. Cosmic rays were identified by comparing pairs of images at each dither position, and were removed by interpolating neighbouring pixels. The spectra were then divided by a flat-field, and sky-subtracted. The spectral trace was measured and corrected to straighten the spectra within the 2D images, and the wavelength solution was derived from the position of the lines within the argon lamp spectrum. The spectrum of $\zeta$~Del~B, and the telluric calibrator, at each nod position was extracted using a nine pixel ($1.35$~arcsec) rectangular aperture extended perpendicular to the dispersion direction. An adjacent aperture was used to construct a sky spectrum to remove any residuals from the sky-subtraction caused by variations in the strength of the sky lines over the course of the observations. Both sets of observations were scaled by the exposure time, such that the spectra were expressed in units of ADU~s$^{-1}$. The six individual spectra of $\zeta$~Del~B were median-combined, divided by the telluric spectrum, and multiplied by a $T_{\rm eff}=19000~$K $\log\left(g\right)=4.5$ {\sc BT-Settl} model atmosphere \citep{Allard:2012fp}, scaled to the 2MASS $K_{\rm S}$ apparent magnitude HIP~104320. No residual spectral features intrinsic to the B3V telluric standard remained within the calibrated spectrum of $\zeta$~Del~B, indicating a good match between the telluric calibrator and the {\sc BT-Settl} model atmosphere.

The final flux-calibrated spectrum of $\zeta$~Del~B is shown in Fig. \ref{fig:GNIRS}. From this spectrum, the apparent magnitudes in the 2MASS~{\it JHK$_{\rm S}$} and MKO~{\it JHK$^{\prime}$K$_{\rm S}$K} photometric systems were computed using the filter transmission curves given in \citet{Cohen:2003gg} and \citet{Tokunaga:2002ex}, respectively. The MKO filter transmission curves were multiplied by the transmission of the atmosphere above Mauna Kea\footnote{http://www.gemini.edu/sciops/telescopes-and-sites/observing-condition-constraints/ir-transmission-spectra}, assuming a precipitable water vapour column of 3~mm. The resulting MKO filter curves are over-plotted on Fig. \ref{fig:GNIRS} for reference. The apparent magnitudes estimated from the flux-calibrated spectrum, and computed absolute magnitudes and corresponding near-infrared colours, are given in Table \ref{tab:zetDel}. The MKO {\it K}$_{\rm S}=15.35\pm0.17$ of $\zeta$~Del~B measured within the MMT/ARIES data set, assuming a negligible colour transformation between the 2MASS and MKO photometric systems for the A-type primary, is consistent with the magnitude derived from the flux-calibrated spectrum.

\section{Properties of $\zeta$ Del B}
\begin{table}
\caption{Properties of the $\zeta$ Del system}
\begin{tabular}{lccc}
Property & Value & Value & Unit\\
\hline
\hline
Identifiers & \multicolumn{2}{c}{$\zeta$ Del, 4 Del, HD 196180}\\
            & \multicolumn{2}{c}{HIP 101589, HR 7871}\\ 
\hline
&$\zeta$~Del~A&$\zeta$~Del~B\\
\hline      
Parallax                           & \multicolumn{2}{c}{$14.82\pm0.23$$^{a}$} & mas\\
Distance                           & \multicolumn{2}{c}{$67.48\pm1.05$$^{a}$} & pc\\
$\mu_{\alpha}$, $\mu_{\delta}$     & \multicolumn{2}{c}{$45.52\pm0.26$, $11.74\pm0.17$$^{a}$} & mas yr$^{-1}$\\
{\it B$_{\rm T}$}                       & $4.768\pm0.014$$^{b}$ & -- & mag \\
{\it V$_{\rm T}$}                        & $4.660\pm0.009$$^{b}$ & -- & mag \\
2MASS {\it J}                          & $4.821\pm0.284$$^{c}$ & $17.17\pm0.09$$^{d}$ & mag \\
2MASS {\it H}                          & $4.484\pm0.033$$^{c}$ & $16.19\pm0.08$$^{d}$ & mag \\
2MASS {\it K$_{\rm S}$}                  & $4.357\pm0.036$$^{c}$ & $15.43\pm0.07$$^{d}$ & mag \\
MKO {\it J}                            & $4.821\pm0.284$$^{e}$ & $17.18\pm0.09$$^{d}$ & mag \\
MKO {\it H}                             & $4.484\pm0.033$$^{e}$ & $16.26\pm0.09$$^{d}$ & mag \\
MKO {\it K$^{\prime}$}                  & $4.357\pm0.036$$^{e}$ & $15.55\pm0.07$$^{d}$ & mag \\
MKO {\it K$_{\rm S}$}                   & $4.357\pm0.036$$^{e}$ & $15.46\pm0.07$$^{d}$ & mag \\
MKO {\it K}                            & $4.357\pm0.036$$^{e}$ & $15.40\pm0.07$$^{d}$ & mag \\
2MASS $M_{\it J}$                        & $0.675\pm0.286$ & $13.03\pm0.10$ & mag \\
2MASS $M_{\it H}$                        & $0.338\pm0.047$ & $12.04\pm0.09$ & mag \\
2MASS $M_{{\it K}_{\rm S}}$              & $0.211\pm0.049$ & $11.28\pm0.08$ & mag \\
{\it B$_{\rm T}$}-{\it V$_{\rm T}$}            & $0.108\pm0.017$ & -- & -- \\
2MASS {\it J}$-${\it H}                        & $0.337\pm0.286$ & $0.98\pm0.13$ & -- \\
2MASS {\it H}$-${\it K$_{\rm S}$}              & $0.127\pm0.049$ & $0.76\pm0.11$ & -- \\
2MASS {\it J}$-${\it K$_{\rm S}$}   & $0.464\pm0.286$ & $1.75\pm0.12$ & -- \\

BC$_{\it V}$                             & $0.088\pm0.012$$^{f}$ & -- & -- \\
BC$_{\it K}$                             & -- & $3.31\pm0.09$$^{g}$ & -- \\
$M_{\rm bol}$                      & $0.523\pm0.037$ & $14.59\pm0.12$ & mag \\
$L_{\star}$                        & $48.63\pm1.66$$^{h}$ & $0.00012\pm0.00001$$^{h}$ & ${\rm L}_{\odot}$\\
$\log (L_{\star}/{\rm L}_{\odot})$ & $1.687\pm0.015$ & $-3.94\pm0.05$ & dex \\
Spectral type                      & A3V$^{i}$ & ${\rm L}5\pm2$$^{j}$ & -- \\
$T_{\rm eff}$                      & 8336$^{k}$ & $1550^{+250}_{-100}$$^{l}$ & K \\
$\log (g)$                         & 3.72$^{k}$ & $5.0^{+0.5}_{-1.0}$ & dex\\
{[Fe/H]}                           & -0.05$^{k}$ & -- & dex\\
Age                                & \multicolumn{2}{c}{$525\pm125$$^{m}$} & Myr \\
Mass                               & $2.5\pm0.2$$^{m}$ & -- & M$_{\odot}$ \\
                                   & -- & $55\pm10$$^{n}$ & $M_{\rm Jup}$ \\
                                   & -- & $40^{+15}_{-5}$$^{o}$ & $M_{\rm Jup}$ \\
Mass ratio                         & \multicolumn{2}{c}{$0.021\pm0.004$$^{n}$} & -- \\
                                   & \multicolumn{2}{c}{$0.015^{+0.006}_{-0.002}$$^{o}$} & -- \\
$\rho$                             & \multicolumn{2}{c}{$13.51\pm0.08$$^{p}$} & arcsec\\
$\theta$                           & \multicolumn{2}{c}{$143.35\pm0.30$$^{p}$} & deg \\
$a_{\rm proj}$                     & \multicolumn{2}{c}{$912\pm15$} & au \\
$a$                                & \multicolumn{2}{c}{$907^{+723}_{-236}$$^{q}$} & au \\
$P$                                & \multicolumn{2}{c}{${\sim}10^4$ $^{r}$} & yr \\
\hline
\label{tab:zetDel}
\end{tabular}

$a$ - \citealp{vanLeeuwen:2007dc}, assumed to be equal for both components, $b$ - \citealp{Hog:2000wk}, $c$ - \citealp{Skrutskie:2006hla}, $d$ - estimated from the flux-calibrated GNIRS spectrum of $\zeta$~Del~B, $e$ - negligible 2MASS-MKO colour transformation for an A3V star, $f$ - interpolated from {\it B}$-${\it V} versus BC$_{\it V}$ table in \citealp{Flower:1996ima}, $g$ - estimated from spectral type -- BC$_{\it K}$ relation in \citealp{Liu:2010cw}, $h$ - calculated from $M_{\rm bol}$, assuming $M_{\rm bol,\odot}=4.74$ \citep{Drilling:2000vo}, $i$ - \citealp{Cowley:1969ew}, $j$ - estimated from comparison with SpeX library (Fig. \ref{fig:SpeX}), $k$ - \citealp{Erspamer:2003hz}, $l$ - estimated from adopted spectral type and equation 3 of \citet{Stephens:2009cc}, $m$ - estimated from \citet{Siess:2000tk} and \citet{Bressan:2012bx} evolutionary models, $n$ - estimated from system age and absolute {\it K$_{\rm S}$} magnitude (Fig. \ref{fig:mass}, top panel), $o$ - estimated from system age and adopted temperature (Fig. \ref{fig:mass}, bottom panel), $p$ - astrometry measured in 2008 Gemini/NIRI observations, $q$ - calculated using the probability density function for the factor $a/a_{\rm proj}$ computed by \citet{Dupuy:2011ip}, assuming a flat eccentricity distribution and no detection bias, $r$ - calculated assuming a face-on, circular orbit.
\end{table}

\begin{figure}
\resizebox{\hsize}{!}{{\includegraphics{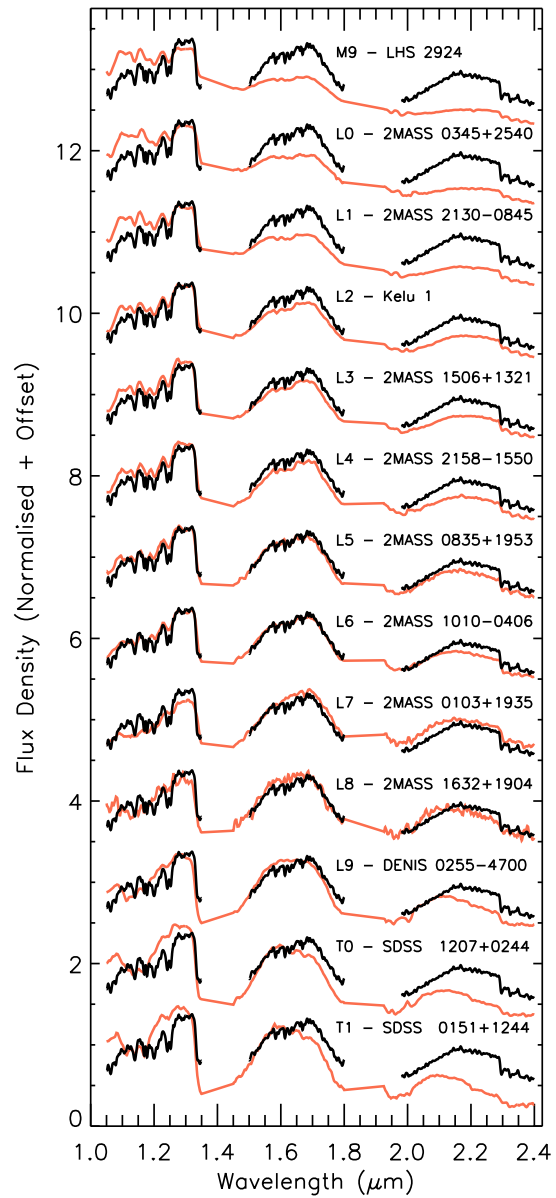}}} 
\caption{The GNIRS spectrum of $\zeta$~Del~B (black curve) is plotted against the M, L, and T-dwarf near-infrared spectral standards from \citet{Kirkpatrick:2010dc} and \citet{Geissler:2011gg} (red curves). The spectral standards have been scaled relative to the spectrum of $\zeta$~Del~B to minimize the reduced $\chi^2$. The observed spectrum of $\zeta$~Del~B is best-fit by the mid-L spectral standards.}
\label{fig:SpeX_std}
\end{figure}
\begin{figure}
\resizebox{\hsize}{!}{{\includegraphics{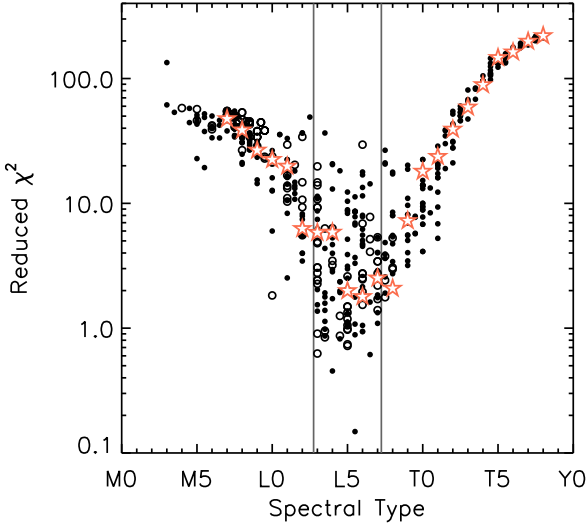}}} 
\caption{The reduced $\chi^2$ of the comparison between the observed GNIRS spectra of $\zeta$~Del~B and the M, L and T-dwarfs within the SpeX library. Near-infrared spectral types were preferentially used; the objects with only an optical spectral type are denoted by open circles. The near-infrared spectral standards of \citet{Kirkpatrick:2010dc} and \citet{Geissler:2011gg} are shown as red stars. The considerable spread in the $\chi^2$ values at a given spectral type can be explained by intrinsic differences between the spectra of brown dwarfs with different surface gravities, and the uncertainty in assigning a specific spectral type to an individual object. The spectrum of $\zeta$~Del~B is most similar to the mid L-dwarfs within the SpeX Library, and as such a spectral type of L$5\pm2$ is adopted (denoted by the vertical lines).}
\label{fig:SpeX}
\end{figure}

\subsection{Spectral type, effective temperature, and surface gravity}
\begin{table}
\caption{Spectral type of $\zeta$~Del~B derived from spectral indices}
\begin{tabular}{cccc}
Index & Value & Spectral type & Reference\\
\hline
\hline   
H$_2$O {\it A} & $0.55\pm0.08$ & L$3.6\pm2.2$ & \cite{McLean:2003hx}\\
H$_2$O {\it B} & $0.61\pm0.14$ & L$5.5\pm3.3$ & \cite{McLean:2003hx}\\
H$_2$O {\it C} & $0.66\pm0.14$ & L$3.1\pm5.6$ & \cite{McLean:2003hx}\\
H$_2$O {\it D} & $0.85\pm0.10$ & L$3.1\pm2.6$ & \cite{McLean:2003hx}\\
H$_2$O   & $1.35\pm0.18$ & L$4.6\pm6.3$ & \cite{Allers:2007ja}\\
H$_2$O {\it J} & $0.78\pm0.13$ & L$4.8\pm3.9$ & \cite{Burgasser:2007fl} \\
H$_2$O {\it H} & $0.72\pm0.10$ & L$5.7\pm3.8$ & \cite{Burgasser:2007fl}\\
CH$_4$ {\it K} & $1.07\pm0.08$ & L$1.6\pm3.9$ & \cite{Burgasser:2007fl}\\
\hline
\label{tab:spectra_indices}
\end{tabular}
\end{table}
\begin{table}
\caption{Measured equivalent widths for $\zeta$~Del~B}
\begin{tabular}{ccc}
Feature & Wavelength & Equivalent width\\
&(\AA)& (\AA) \\
\hline
\hline   
Na {\sc i} & 11396 & $11.3\pm0.5$\\
K {\sc i} & 11692 & $10.8\pm0.4$\\
K {\sc i} & 11778 & $11.9\pm0.4$\\
K {\sc i} & 12437 & $6.8\pm0.4$\\
K {\sc i} & 12529 & $9.4\pm0.7$\\
\hline
\label{tab:equiv_widths}
\end{tabular}
\end{table}
The spectral type of $\zeta$~Del~B was estimated through a comparison with empirical spectra of field brown dwarfs within the SpeX Prism Spectral Library\footnote{http://www.browndwarfs.org/spexprism}. The GNIRS spectrum of $\zeta$~Del~B and the SpeX spectra were smoothed by convolution with a 50\AA~Gaussian, and interpolated to the same wavelength scale. For each object within the SpeX library, the spectrum was scaled to minimize the reduced $\chi^2$ when compared with $\zeta$~Del~B, calculated over the wavelength ranges $1.05-1.35$~\micron, $1.45-1.80$~\micron, and $1.98-2.40$~\micron. In this case, the $\chi^2$ statistic is equivalent to the $G$ statistic, defined by \citet{Cushing:2008kb}, as the weightings are uniform across the entire wavelength range of the spectrum of $\zeta$~Del~B. Fig. \ref{fig:SpeX_std} shows the spectrum of $\zeta$~Del~B compared with the M, L and T-dwarf near-infrared spectral standards from \citet{Kirkpatrick:2010dc} and \citet{Geissler:2011gg}, which are all available within the SpeX library. Fig. \ref{fig:SpeX} shows the reduced $\chi^2$ as a function of spectral type for all of the objects within the SpeX library, with the spectral standards highlighted, from which a spectral type of L$5\pm2$ was adopted for $\zeta$~Del~B. The mid-L spectral classification is consistent with the spectral type estimated from various spectral indices defined in the literature, given in Table \ref{tab:spectra_indices} \citep{McLean:2003hx,Allers:2007ja,Burgasser:2007fl}. This range of spectral types corresponds to an effective temperature of $1550^{+250}_{-100}$~K, using the spectral type-temperature relations of \citet{Stephens:2009cc} derived for field brown dwarfs. The spectrum of $\zeta$~Del~B was also compared with {\sc BT-Settl} synthetic spectra \citep{Allard:2012fp}, with a best fitting effective temperature of $T_{\rm eff}=1650\pm200$~K and surface gravity of $\log \left(g\right)=5.0^{+0.5}_{-1.0}$, based on a $\chi^2$ minimization of the observed and model spectra.

The observed spectrum does not exhibit the strong triangular shape of the {\it H}-band continuum seen in the lowest-surface gravity brown dwarfs (e.g. \citealp{Lucas:2001ed,Allers:2007ja}), consistent with the higher surface gravity estimated for $\zeta$~Del~B from the comparison with the model spectra. The shape of the {\it H} band peak can be quantified using the {\it H}-cont index defined by \citet{Allers:2013hk}. While the {\it H}-cont index for $\zeta$~Del~B of 0.92 would suggest a lower surface gravity than field objects, \citet{Allers:2013hk} note that numerous older dusty field brown dwarfs also exhibit similar behaviour, and caution against using this index alone to diagnose low surface gravity. The remaining \citet{Allers:2013hk} indices which could be measured using the GNIRS data, FeH$_{\it J}= 1.2$ and K~I$_{\it J}= 1.2$, are consistent with the field population, and suggest a high surface gravity for $\zeta$~Del~B.

Individual spectral features are also diagnostic of surface gravity. The strength of the sodium (Na {\sc i}) line at 1.1396~$\micron$ and the potassium (K {\sc i}) doublets at 1.174 and 1.248~$\micron$, indicated in Fig. \ref{fig:GNIRS}, are observed to depend strongly on the spectral type and surface gravity of an object \citep{Gorlova:2003cs}, being significantly weaker within the spectra of the youngest objects of a given spectral type \citep{Allers:2013hk}. The equivalent widths of these lines, which were calculated using the method described in \citet{McLean:2003hx}, are given in Table \ref{tab:equiv_widths}. The strength of these lines within the spectrum of $\zeta$~Del~B suggest a surface gravity more similar to those found for field brown dwarfs (e.g. \citealp{McLean:2007ir}), than found for younger objects (e.g. \citealp{Bonnefoy:2014dh}). The presence of a deep $^{12}$CO 2-0 bandhead at 2.29~$\micron$, indicated in Fig. \ref{fig:GNIRS}, is a potential indicator of low surface gravity (e.g. \citealp{Cushing:2005ed}), based on the depth of the bandhead within the spectra of M-type giants relative to M-dwarfs of the same spectral type \citep{Kleinmann:1986gm}. Without an empirical relation between the strength of this absorption and surface gravity for a given brown dwarf spectral type, the significance of the deep absorption seen in the spectrum of $\zeta$~Del~B cannot be quantified.

\subsection{Bolometric luminosity and mass}
\begin{figure}
\resizebox{\hsize}{!}{{\includegraphics{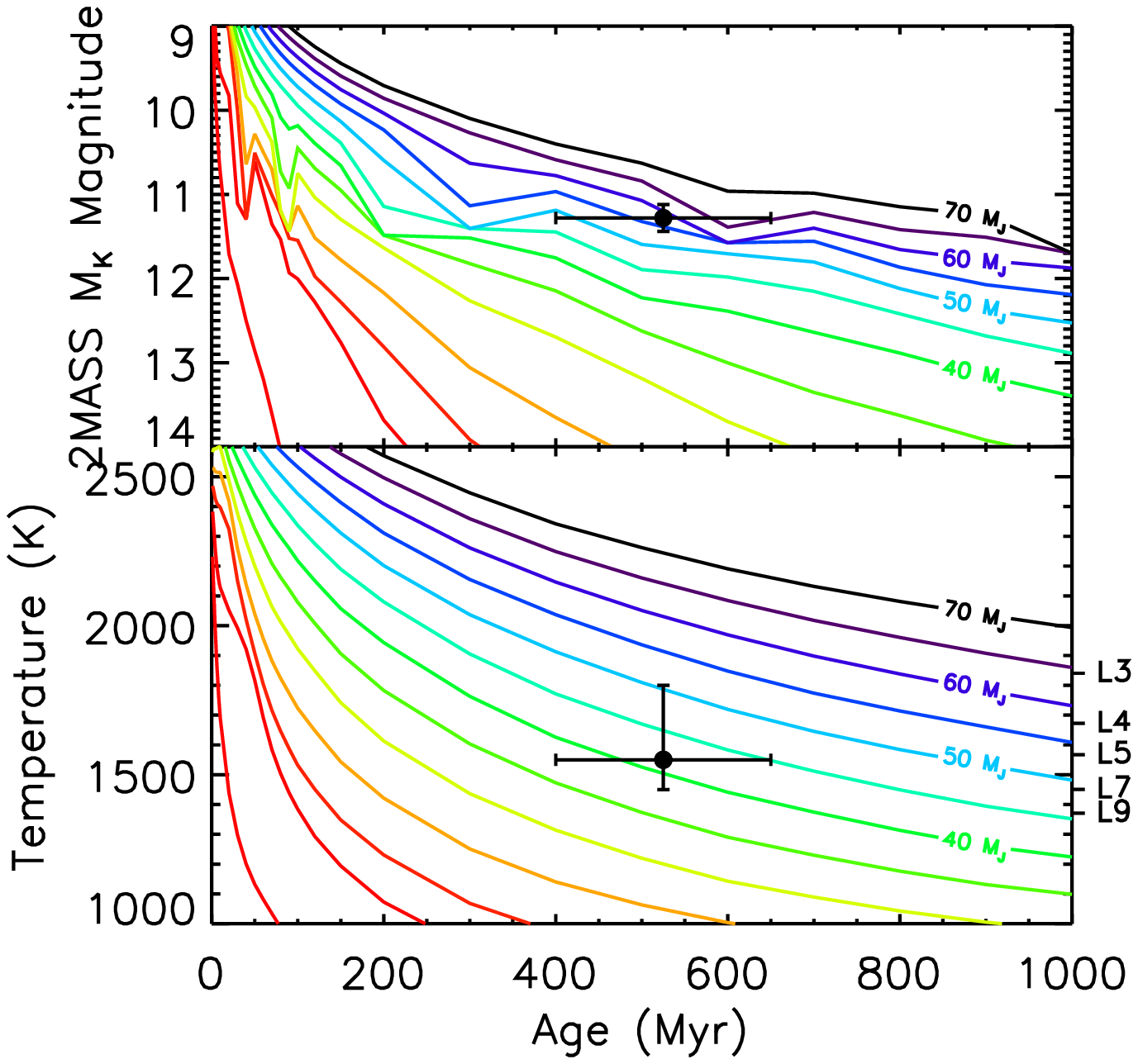}}} 
\caption{Atmospheric/evolutionary models showing the decline in absolute 2MASS {\it K$_{\rm S}$} magnitude (top panel) and temperature (bottom panel) as a function of age for substellar objects between 10 and 70~$M_{\rm Jup}$ \citep{Allard:2012fp}. Using the age of $\zeta$~Del~A and the absolute magnitude of $\zeta$~Del~B of $M_{\it K_{\rm S}} = 11.14\pm0.17$, we estimate a mass of $55\pm10$~$M_{\rm Jup}$ (top panel). Using the adopted temperature of $1550^{+250}_{-100}$~K, derived from the spectral type-temperature relations given in \citet{Stephens:2009cc}, yields a lower mass estimate of $40^{+15}_{-5}$~$M_{\rm Jup}$ (bottom panel).}
\label{fig:mass}
\end{figure}
In order to determine the bolometric luminosity of $\zeta$~Del~B, a {\it K}-band bolometric correction of BC$_{\it K}=3.31\pm0.09$ was estimated using the adopted spectral type and the spectral type-BC$_{\it K}$ relation for field brown dwarfs presented in \citet{Liu:2010cw}. Applying this correction to the measured {\it K}-band magnitude leads to a bolometric magnitude of $M_{\rm bol}=14.59\pm0.12$ and a bolometric luminosity of $\log\left(L/{\rm L}_{\odot}\right)=-3.94\pm0.05$ for $\zeta$~Del~B, assuming $M_{\rm bol,\odot}=4.74$ \citep{Drilling:2000vo}. Using evolutionary models at an age of $525\pm125$~Myr \citep{Baraffe:2002di}, this bolometric luminosity corresponds to an effective temperature of ${\sim}1950\pm100$~K, higher than the values estimated from both the adopted spectral type and the best-fit model atmosphere. A similar discrepancy is also seen for the 800~Myr brown dwarf binary HD~130948 BC \citep{Dupuy:2009jq} and the 100~Myr brown dwarf CD-35~2722~B \citep{Wahhaj:2011by}, where the effective temperatures derived from the evolutionary models using the age and bolometric luminosity are ${\sim}100-300$~K hotter than predicted from the spectral type.

The {\sc BT-Settl} atmospheric models \citep{Allard:2012fp} and the \citet{Baraffe:2002di} evolutionary models were used to provide an estimate of the mass of $\zeta$~Del~B, exploiting the strong dependence on age of the mass of substellar objects of a given luminosity or temperature. Using the measured {\it K}-band magnitude for $\zeta$~Del~B, and assuming an age of the system of $525\pm125$~Myr, the evolutionary models yield a mass of $55\pm10$~$M_{\rm Jup}$ (Fig. \ref{fig:mass}, top panel). Alternatively, using the effective temperature derived from the adopted spectral type leads to a lower mass estimate of $40^{+15}_{-5}$~$M_{\rm Jup}$ (Fig. \ref{fig:mass}, bottom panel). Combining these mass estimates with the mass of the primary given previously, the companion-to-primary mass ratio of the $\zeta$~Del system was estimated to be between $0.023\pm0.004$ using the {\it K}-band magnitude and $0.015^{+0.006}_{-0.002}$ using the adopted effective temperature.

\subsection{Probability of chance superposition}
Using the L$5\pm2$ spectral type for $\zeta$~Del~B and estimates for the surface density of mid-L dwarfs, an estimate of the probability of a chance superposition between $\zeta$~Del~A and a unassociated foreground or background brown dwarf can be calculated. The surface density of L3-L7 brown dwarfs within a simulated magnitude-limited (${\it K}<16$) survey, assuming a log-normal mass distribution \citep{Chabrier:2002hm}, was estimated to be $1.42 \times 10^{-2}$~deg$^{-2}$ \citep{Burgasser:2007fl}. Assuming brown dwarfs are isotropically distributed throughout the sky, this surface density would correspond to a probability of detecting a brown dwarf of spectral type between L3 and L7 within a radius of $13.5$~arcsec from a random location on the sky of $P\approx2\times 10^{-3}$. This is an upper limit to the true probability as the surface density includes contributions from all L3-L7 brown dwarfs along the line-of-sight until the magnitude limit of ${\it K}<16$ is reached at $D\approx 100$~pc. As the spectral type of $\zeta$~Del~B provides a reasonable limit on its distance ($25\la D{\rm[pc]} \la 110$), the object can be constrained to a volume of space defined by a truncated cone with an inner and outer radii of $1.6\times10^{-3}$ and $7.2\times10^{-3}$~pc, respectively, and a height of $85$~pc. Combining this volume with the number density of L3-L7 brown dwarfs, $3.87\times10^{-3}$~pc$^{-3}$ \citep{Burgasser:2007fl}, yields a probability of detecting an object like $\zeta$~Del~B within a radius of $13.5$~arcsec from a random location on the sky of $P\approx 2\times10^{-5}$.

\subsection{Semimajor axis and orbital period}
The angular separation of $13.51\pm0.08$~arcsec between $\zeta$~Del~A and B measured within the 2008 Gemini/NIRI dataset was converted to a projected separation of $a_{\rm proj}=912\pm15$~au using the {\it Hipparcos} parallax for $\zeta$~Del~A ($\pi=14.82\pm0.23$~mas; \citealp{vanLeeuwen:2007dc}). Due to the large number of orbital orientations and viewing geometries, the conversion between projected separation as seen on the sky into a semimajor axis $\left(a\right)$ is non-trivial. Using the probability density function calculated for $a/a_{\rm proj}$ by \citet{Dupuy:2011ip}, in the case of a flat eccentricity distribution with no discovery bias, the most probable semimajor axis of $\zeta$~Del~B was estimated via a Monte Carlo method as $907^{+723}_{-236}$~au. An order-of-magnitude estimate of the orbital period of ${\sim}10^4$~yrs was made using the most probable semimajor axis, assuming a face-on circular orbit, and the masses of each component given in Table \ref{tab:zetDel}.

\subsection{Comparison with known substellar companions}
\begin{figure}
\resizebox{\hsize}{!}{{\includegraphics{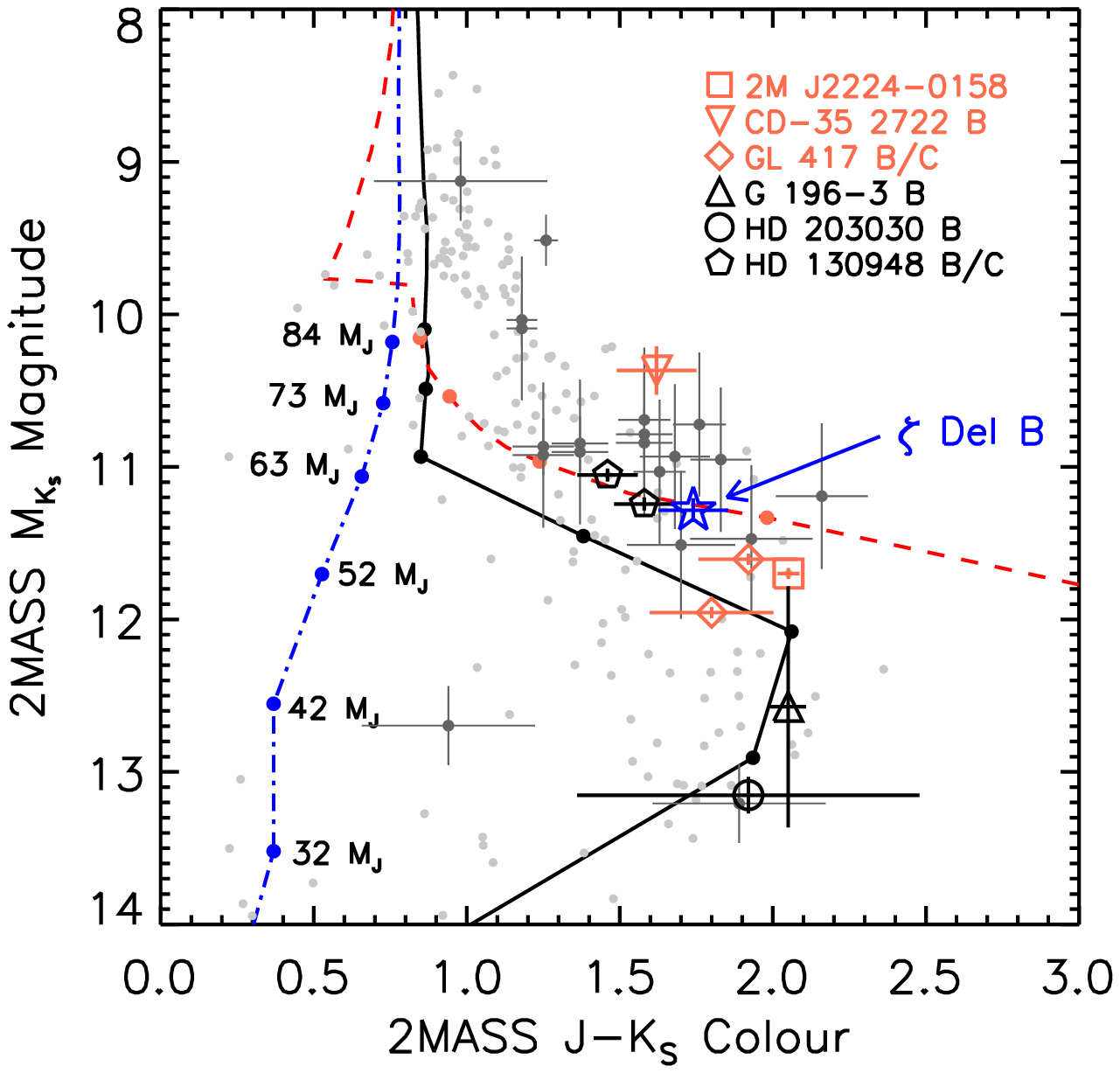}}} 
\caption{CMD in the 2MASS photometric system \citep{Skrutskie:2006hla} showing the location of $\zeta$~Del~B (blue open star) relative to field brown dwarfs (light grey circles; \citealp{Dupuy:2012bp}) and known Hyades brown dwarfs (filled circles; \citealp{Reid:1993tv,Bouvier:2008kf,Hogan:2008ha}). Known substellar companions with ages of 100-1000~Myr drawn from the compilation of \citet{Zuckerman:2009gc}, and the recently-discovered companion CD-35~2722~B \citep{Wahhaj:2011by}, are plotted for reference (open black symbols). The dusty field L4.5 dwarf 2MASS~J22244381-0158521 \citep{Cushing:2005ed} is also highlighted (open red square). The literature {\it J} and {\it K}-band photometry for CD-35~2722~B and HD~130948~BC, and {\it J}-band photometry for HD~203030~B, were measured in the MKO photometric system; no 2MASS magnitudes for these objects were found within the literature. For mid L-dwarfs, the colour correction between these photometric systems is expected to be $|\Delta${\it J}$-${\it K}$|\leq0.2$~mags (e.g. \citealp{Stephens:2004br}). Theoretical 500~Myr isochrones from the \textsc{BT-Settl} (black solid curve; \citealp{Allard:2012fp}), \textsc{AMES-Cond} (blue dot-dashed curve; \citealp{Baraffe:2003bj}), and \textsc{AMES-Dusty} (red dashed curve; \citealp{Chabrier:2000hq}) model grids are shown for comparison.}
\label{fig:CMD}
\end{figure}
\begin{figure*}
\resizebox{\hsize}{!}{{\includegraphics{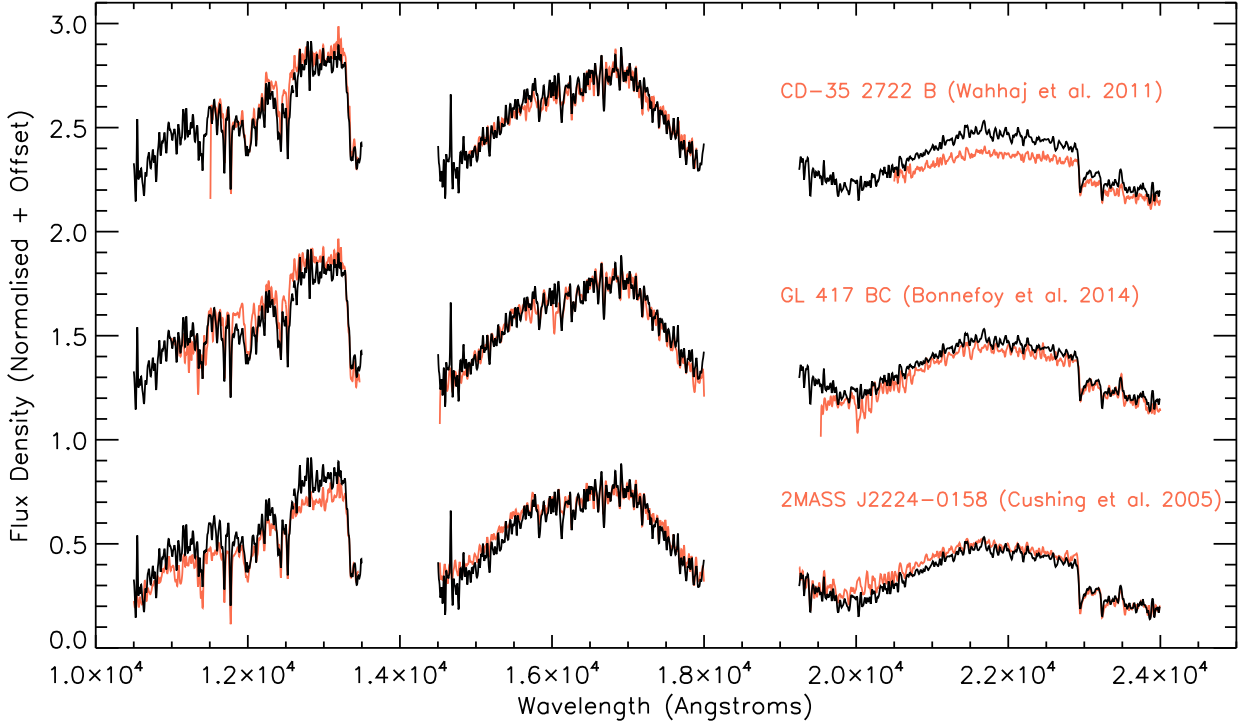}}} 
\caption{The spectrum of $\zeta$~Del~B (black curve) compared to three of the most analogous objects for which intermediate-resolution spectra exist; CD-35~2722~B (top red curve; \citealp{Wahhaj:2011by}), the blended spectrum of Gl~417~BC (middle red curve; \citealp{Bonnefoy:2014dh}), and 2MASS~J2224-0158 (bottom red curve; \citealp{Cushing:2005ed}). Each spectrum is normalized to the flux measured in the {\it H}-band between 1.65 and 1.75~$\micron$.}
\label{fig:spectra_comparison}
\end{figure*}

The position of $\zeta$~Del~B on a $M_{\it K}$ versus {\it J}$-${\it K$_{\rm S}$} CMD, shown in Fig. \ref{fig:CMD}, was compared to known intermediate-age ($100-1000$~Myr) substellar companions, Hyades brown dwarfs \citep{Reid:1993tv,Bouvier:2008kf,Hogan:2008ha}, and older field brown dwarfs \citep{Dupuy:2012bp}. The intermediate-age comparison sample of 100-1000~Myr substellar companions was drawn from the compilation of \citet{Zuckerman:2009gc}, complemented with the recent discovery of the ${\sim}100$~Myr L4 dwarf CD-35~2722~B \citep{Wahhaj:2011by}. Several of the included objects do not have a parallax measurement within the literature - for G~196-3 a photometric distance of $11\pm4$~pc was used \citep{Shkolnik:2012cs}, and for the Hyades brown dwarfs without distance estimates a value of $47.5\pm3.6$~pc was used \citep{McArthur:2011dk}. For the three objects without stated photometric uncertainties, a value of $\sigma_m=0.2$~mag was assumed. $\zeta$~Del~B is at a similar location on the CMD to each of the components of the binary brown dwarfs Gl~417~BC (80-300~Myr, L4.5+L6; \citealp{Zuckerman:2009gc}) and HD~130948 BC (800~Myr, L4+L4; \citealp{Dupuy:2009jq}), and has a similar colour to the recently-discovered companion to CD-35~2722 (100~Myr, L4; \citealp{Wahhaj:2011by}). The position of $\zeta$~Del~B on the CMD is also consistent with the predicted location of a $\sim 50$~$M_{\rm Jup}$ object at 500~Myr within the {\sc AMES-Dusty} model grid, intermediate to the two mass estimates for $\zeta$~Del~B given in Table~\ref{tab:zetDel}.

The spectrum of $\zeta$~Del~B is compared in Fig. \ref{fig:spectra_comparison} with the most analogous objects for which similar-resolution spectra were available: CD-35~2722~B \citep{Wahhaj:2011by}, 2MASS~J22244381-0158521 (2MASS~J2224-0158; \citealp{Cushing:2005ed}), and the blended spectrum of the brown dwarf binary Gl~417~BC \citep{Bonnefoy:2014dh}. While a good match is seen between these three objects and $\zeta$~Del~B when each bandpass is fitted independently, $\zeta$~Del~B appears redder than CD-35~2722~B and bluer than 2MASS~J2224-0158 when the full flux-calibrated {\it JHK} spectra are compared, consistent with their relative positions on the CMD (Fig. \ref{fig:CMD}).

\section{Wide substellar companion formation}
\begin{figure}
\resizebox{\hsize}{!}{{\includegraphics{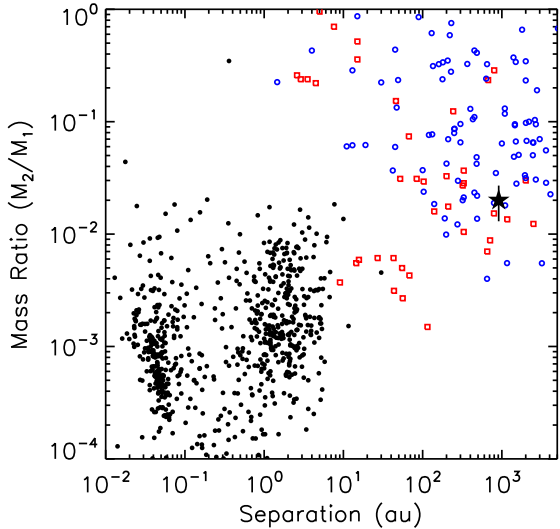}}} 
\caption{Mass ratio as a function of separation for brown dwarf companions given in \citet{Zuckerman:2009gc} and \citet{Deacon:2014tf} (blue open circles), imaged companions listed in exoplanets.eu (red open squares), and companions detected by radial velocity and transiting in exoplanets.eu (black points). $\zeta$~Del~B (black filled star) is among the most widely separated, lowest mass ratio companions imaged to date.}
\label{fig:q-ratio}
\end{figure}

$\zeta$~Del~B is among the most widely-separated and lowest mass ratio companions resolved around a main-sequence star to-date, as shown in Fig. \ref{fig:q-ratio}. Significant uncertainty exists in the formation history of such objects. Intermediate to the lowest-mass stars and massive directly-imaged planetary-mass companions, a number of formation theories have been suggested for these objects, ranging from formation in a large circumstellar disc, or through the fragmentation of a pre-stellar core, to the gravitational capture by the massive primary within a dynamical star-forming environment. While disc fragmentation models predict the formation of substellar companions around A-type primaries (e.g. \citealp{Kratter:2010gf}), the formation of $\zeta$~Del~B {\it in situ} would require an unusually massive circumstellar disc for sufficient mass to be available at such separations. The time-scale for formation through core accretion is also prohibitively long at large separations (e.g. \citealp{Pollack:1996jp}), requiring the core to form before the gas disc dissipates. These formation scenarios cannot be excluded, however, as $\zeta$~Del~B may have migrated outward prior to the dissipation of the disc (e.g. \citealp{Vorobyov:2013dl}), or dynamical interaction with unseen interior companion may have dynamically scattered it (e.g. \citealp{Jiang:2004dd,Stamatellos:2009kg}) on to a potentially unstable orbit (e.g. \citealp{Veras:2009br}).

Formation through the fragmentation of a pre-stellar core is predicted to form companions with separations of the order of $10^3$~au \citep{Bate:1995wm}, with mass ratios ranging from the stellar to substellar regimes \citep{Bonnell:1992ep}. Large-scale simulations of star formation within clusters are able to produce low-mass ratio companions at wide separations, although such models are limited in terms of the overall size of the cluster and, by extension, the number of high-mass stars produced \citep{Bate:2009br,Bate:2011hy}. An alternative scenario is that $\zeta$~Del~B formed independent of $\zeta$~Del~A, and was dynamically captured, either through a three-body interaction in which a natal companion was ejected (e.g. \citealp{Bonnell:2001tb}), or through disc-assisted capture where the angular momentum of the substellar impactor was dissipated by a large circumstellar disc (e.g. \citealp{Moeckel:2007hm}). However, given the predicted low efficiency of binary formation through capture \citep{Clarke:1991wg}, this scenario is unlikely.

\section{Summary}
We have presented the discovery of a wide substellar companion to the A3V star $\zeta$~Delphini, augmenting the small number of examples of such companions to early-type stars currently reported within the literature. Based on our AO images obtained in 2008, $\zeta$~Del~B is at a projected separation of $a_{\rm proj}=912\pm15$~au from $\zeta$~Del~A, with a most probable semi-major axis of $a=907^{+723}_{-236}$~au, assuming a flat eccentricity distribution \citep{Dupuy:2011ip}. Based on the position of $\zeta$~Del~A on the temperature-surface gravity and CMDS, we estimate a system age of $525\pm125$~Myr. Comparing $\zeta$~Del~B with evolutionary models, we estimate a mass of $55\pm10$~$M_{\rm Jup}$ using bolometric luminosity estimated from the {\it K$_{\rm S}$} magnitude, and $40_{-5}^{+15}$~$M_{\rm Jup}$ using the adopted spectral type and empirical spectral type-temperature relations, corresponding to a companion mass ratio of $q=0.02\pm0.01$. Future spectroscopic measurements of the C/O ratio of $\zeta$~Del~B, and other wide substellar companions, may provide an observational diagnostic of their formation mechanism (e.g. \citealp{Konopacky:2013jv}). The continuation of large-scale surveys for such objects (e.g. \citealp{Faherty:2010gt,Deacon:2014tf,Naud:2014jx}) is essential to develop empirical comparisons for theoretical formation models. The frequency and properties of such objects at wide separations ($\sim10^3$~au) will also provide important context for the expected yield of substellar companions resolved in upcoming extreme-AO systems such as GPI \citep{Macintosh:2014wa} and SPHERE \citep{Beuzit:2008gt} at separations of between $1-100$~au to nearby main-sequence stars.

\section*{Acknowledgements} The authors wish to express their gratitude for the constructive comments received from the referee. The authors gratefully acknowledge several sources of funding. RJDR acknowledges financial support from the Science and Technology Facilities Council (ST/H002707/1). KWD is supported by an NSF Graduate Research Fellowship (DGE-1311230). Portions of this work were performed under the auspices of the U.S. Department of Energy by Lawrence Livermore National Laboratory in part under Contract W-7405-Eng-48 and in part under Contract DE-AC52-07NA27344, and also supported in part by the NSF Science and Technology CfAO, managed by the UC Santa Cruz under cooperative agreement AST 98-76783. This work was supported, through JRG, in part by University of California Lab Research Programme 09-LR-118057-GRAJ and NSF grant AST-0909188. Based on observations obtained at the Canada-France-Hawaii Telescope (CFHT) which is operated by the National Research Council of Canada, the Institut National des Sciences de l'Univers of the Centre National de la Recherche Scientifique of France, and the University of Hawaii. Based on observations obtained at the Gemini Observatory, which is operated by the Association of Universities for Research in Astronomy, Inc., under a cooperative agreement  with the NSF on behalf of the Gemini partnership: the National Science Foundation (United States), the Science and Technology Facilities Council (United Kingdom), the National Research Council (Canada), CONICYT (Chile), the Australian Research Council (Australia),  Minist\'{e}rio da Ci\^{e}ncia e Tecnologia (Brazil) and Ministerio de Ciencia, Tecnolog\'{i}a e Innovaci\'{o}n Productiva (Argentina). This research has made use of the SIMBAD data base and the VizieR catalogue access tool, operated at CDS, Strasbourg, France. This research has benefited from the SpeX Prism Spectral Libraries, maintained by Adam Burgasser at http://www.browndwarfs.org/spexprism. This publication makes use of data products from the Two Micron All Sky Survey, which is a joint project of the University of Massachusetts and the Infrared Processing and Analysis Center/California Institute of Technology, funded by the National Aeronautics and Space Administration and the National Science Foundation. This publication makes use of data products from the Wide-field Infrared Survey Explorer, which is a joint project of the University of California, Los Angeles, and the Jet Propulsion Laboratory/California Institute of Technology, funded by the National Aeronautics and Space Administration. This research used the facilities of the Canadian Astronomy Data Centre operated by the National Research Council of Canada with the support of the Canadian Space Agency.

\bibliographystyle{mn2e}
\bibliography{paper}
\label{lastpage}
\end{document}